\documentclass[%
aip,
jap,%
sd,%
amsmath,amssymb,
preprint,%
]{revtex4-1}

\usepackage{graphicx}
\usepackage{dcolumn}
\usepackage{bm}

\begin{document}


This draft was prepared using \textit{LaTeX} template provided by Journal of Applied Physics

\title{Droplet spreading: Role of characteristic length scales and modified disjoining pressure}

\author{Debanik Bhattacharjee}
\author{Hadi Nazaripoor}
\author{Mohtada Sadrzadeh}
\email{sadrzade@ualberta.ca}
\affiliation{Advanced Water Research Lab, Department of Mechanical Engineering, University of Alberta, AB-T6G 2R3}

\date{\today}

\begin{abstract}
The numerical prediction of droplet base radius and contact angle depends on the choice of characteristic radius and height of droplet. In this study, a developed model on the basis of lubrication approximation is used to investigate the effect of characteristic parameters on the spreading of droplets over substrates. Based on two main stages of spreading (initial and equilibrium), characteristic radius and height were first evaluated. The model predictions were then compared with experimental results in literature for both impermeable and permeable substrates. The study provides evidence that the choice of characteristic length scales based on either stage enables accurate prediction of the droplet base radius and contact angle. In addition, a new scaling relation for the predictions of the numerical disjoining pressure, with an error of $\pm$5\%, is proposed.
\end{abstract}

\maketitle

\section{\label{sec:level1}INTRODUCTION}

Droplet spreading is ubiquitous in nature and has been exploited in many applications such as microfluidic devices, \cite{Choi2012} printing processes, \cite{Kuang2014} spray cooling, \cite{Mousa2009} and tissue assays. \cite{Jia2003} The spreading of a liquid on a substrate is dictated by physical parameters,\cite{Benintendi1999,Kovalchuk2014,Erbil1997} topography, \cite{McHale2009,Courbin2007,Savva2012} and chemical heterogeneities. \cite{Modaressi2002,Bliznyuk2009,Herminghaus2008} The spreading behaviour of a perfectly wetting liquid on a smooth, impermeable substrate is governed by Tanner's law  that describes the time evolution of droplet radius ($\sim$ time$^{1/10}$) and droplet height ($\sim$ time$^{-1/5}$).\cite{Tanner1979} Permeability of the substrate is another pertinent feature which affects the spreading of the droplet. \cite{Frank2012,Acton2001} Permeation plays a significant role in the latter stages of spreading which is marked by retraction of the contact line and consequently, complete imbibition of the droplet volume. The characteristics of a perfectly wetting liquid spreading on a dry, permeable substrate fall on two universal curves.\cite{Starov2004} However, no such universal behavior has been observed for partially wetting liquids on permeable substrates. 

Droplet spreading on a substrate is a classical example of a moving contact line problem. Numerical techniques have been found as a versatile tool to track the interface height in droplet spreading studies. Some of the most commonly used techniques include Molecular Dynamics simulation,\cite{He2003,De1999,Song2013} Lattice Boltzmann method,\cite{Rai2000,Attar2009,Dupuis2004} and lubrication based approach.\cite{Ehrhard1991,Espin2015,Benintendi1999,Zadrazil2006} The scaling factors in the lubrication based approach are radius and height of the droplet. The tri-phase contact line is prone to shear-stress singularity due to the no-slip condition which has been addressed through a number of approaches using lubrication approximation. The most common approach is to use a slip model between the liquid and the solid to relieve the stress singularity. Using this model results in a nonlinear equation which can be solved either asymptotically \cite{Greenspan1978,Cox1986,Haley1991,Bostwick2013,Ehrhard1991} or numerically. \cite{Benintendi1999,Smith1995} A constitutive relation describes the contact line velocity in terms of the apparent contact angle. Although slip model provides accurate results, its main shortcoming, however, lies in the accurate prediction of the proportionality constant and the spreading exponent in the constitutive relation since the spreading exponent varies for different liquid-substrate combinations.\cite{McHale2009}

Another approach to remove the contact line singularity is precursor film assumption. The contact line region is described by disjoining pressure which consists of attractive and repulsive components. \cite{Gomba2010,Zadrazil2006,Espin2015} The precursor film or ``primary film'' was first observed for the water and acetic acid drops spreading on glass and steel substrates.\cite{Hardy1919} The thickness of the precursor film was estimated to be few hundreds angstrom using ellipsiometry and interferometry techniques, \cite{Bascom1964} scanning electron microscopy,\cite{Radigan1974} and polarised reflection microscopy.\cite{Ghiradella1975} The main challenge of this approach is the sensitivity of the equilibrium contact angle to the disjoining pressure parameter. 

Identifying a proper characteristic radius and height is the first step towards development of a predictive model for droplet spreading over substrates. The characteristic values are typically selected at the onset of equilibrium from experimental data.\cite{Espin2015} A possible explanation for selecting the equilibrium characteristic values might be related to the experimental measurement and easier evaluation of equilibrium values. However, these studies would have been more interesting if the sensitivity of the results to the selection of other characteristic values is included. For a partially wetting liquid, the equilibrium contact angle has a finite non-zero value and the equilibrium stage is interpreted to be controlled by disjoining pressure. The disjoining parameter, $\Pi=A\Big[(\frac{b}{h})^n- (\frac{b}{h})^m\Big]$ is a combination of repulsive (first term) and attractive (second term) components acting when the droplet height is small and comparable to the precursor film thickness, $b$, $(n,m)=(3,2)$, and $A$ is the disjoining pressure parameter. It has been shown that $A$ is inversely proportional to the precursor film thickness. It also depends on the equilibrium contact angle and $(n,m)$.\cite{Schwartz1998} However, none of the proposed relations in the literature provide a viable $A$ which leads to a better agreement between numerical prediction and experimentally measured contact angle.

The present paper is aimed at investigating the effect of characteristic length scales (radius and height) on the spreading of partially wetting droplet over both impermeable and permeable substrates. A mathematical model is used to describe the contact line region through precursor film assumption.\cite{Espin2015} Using the no-slip condition at the substrate, a nonlinear partial differential equation is derived which describes the spatio-temporal evolution of droplet height based on lubrication approximation. A disjoining pressure model is employed to remove contact line singularity.\cite{Gomba2010, Zadrazil2006, Alleborn2004} For imbibition, Poiseuille flow driven by excess liquid pressure is assumed in each unconnected vertical pore which essentially removes the need for thickness-dependent permeability.\cite{Davis1999} Normalising factors are defined based on the initial and equilibrium stages of droplet spreading. Using reported experimental data,\cite{Ruijter1997,Kumar2006,Wang2016} the aim is to investigate whether the prediction of droplet base radius and contact angle is influenced by the stage-dependent characteristic radius and height. In addition, the numerical disjoining pressure parameter is predicted based on its theoretical value through different stages of spreading. 

Section IIA includes details about the governing equations used. Section IIB presents boundary conditions needed for solving the resulting partial differential equation. In section IIC, the variables which are used to non-dimensionalise the governing equations and boundary conditions, are presented. The derivation of the time evolution equation is discussed in section IID along with the numerical scheme utilised to solve the equation. The normalising factors are defined in section IIE. In section III, a comparison of the numerical predictions with the experimental data is presented.\cite{Ruijter1997,Kumar2006,Wang2016} In section IV, a new scaling to bridge the gap between the theoretical and model value of the disjoining pressure parameter is proposed for both impermeable and permeable substrates. All findings are summarised in section V.

\section{\label{sec:level2}MATHEMATICAL FORMULATION}

\begin{figure}
\includegraphics[scale=0.3]{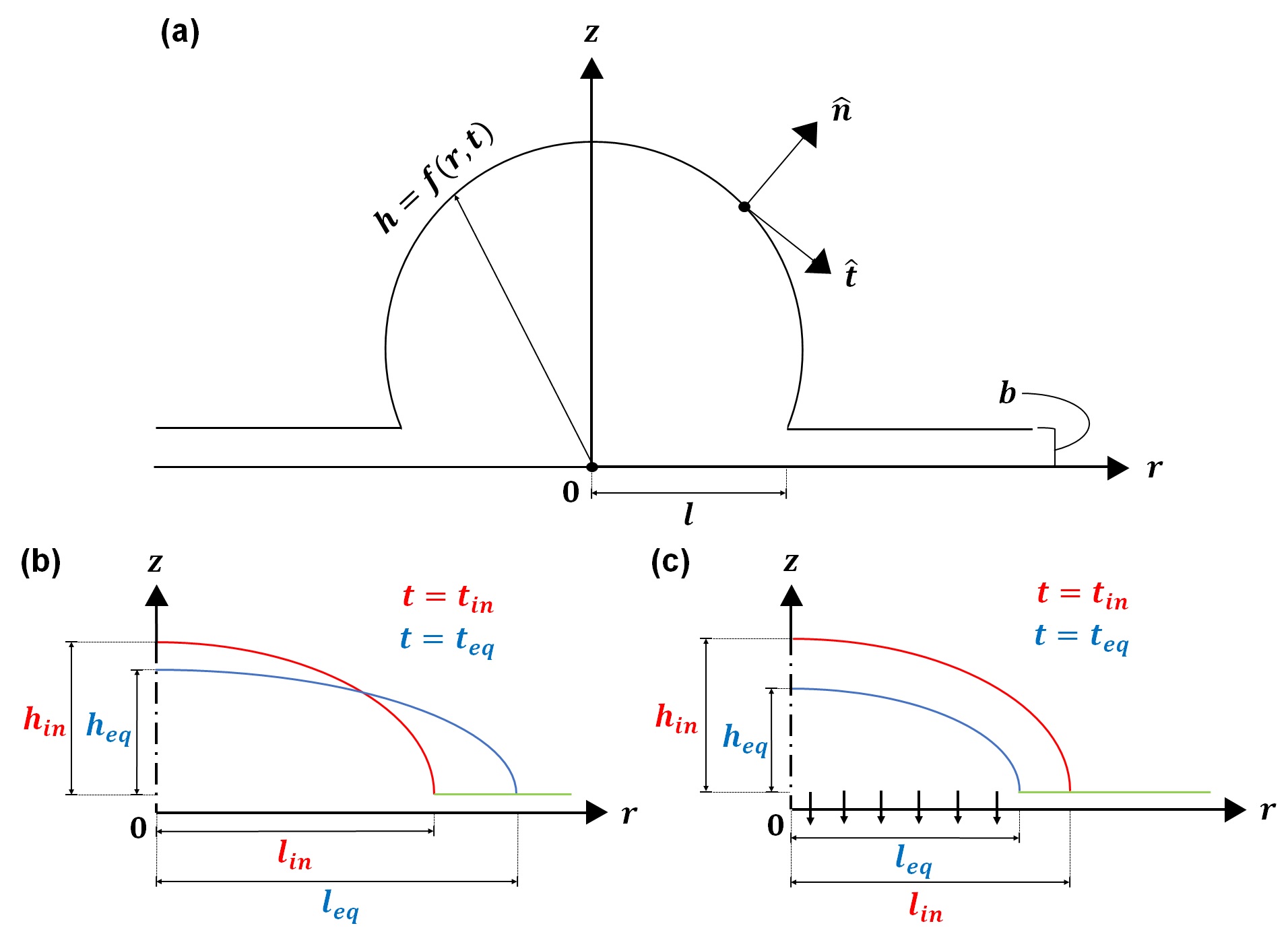}
\caption{(a) Geometrical configuration of an axi-symmetric droplet and different stages of droplet spreading on (b) an impermeable substrate and (c) a permeable substrate. The subscript 'in' and 'eq' denote the initial and equilibrium value of a quantity, respectively.}
\label{fig:one}
\end{figure}

An axi-symmetric liquid droplet is considered to spread on a smooth, horizontal substrate (Fig.~\ref{fig:one}(a)). The dynamics of droplet spreading depends on the nature of the substrate and is different for impermeable and permeable substrate, as illustrated in Fig.~\ref{fig:one}(b) and Fig.~\ref{fig:one}(c). The droplet is assumed to be an incompressible, Newtonian liquid with constant density $\rho$, dynamic viscosity, $\mu$ and surface tension, $\sigma$. Since the droplet is assumed to be axi-symmetric in nature, the governing equations are considered in cylindrical coordinates. 

\subsection{\label{sec:levela}GOVERNING EQUATIONS}

To describe the droplet spreading, pressure, $p$ and velocity field, $\textbf{v}=(u,w)$ is introduced where $u$ and $w$ denote the velocity in radial and vertical directions respectively. The continuity equation and Navier-Stokes equations are as follows:
\begin{equation}\label{eq:1}
\nabla.\textbf{v} = 0\\
\end{equation}
\begin{equation}\label{eq:2}
\rho(\frac{\partial \textbf{v}}{\partial t} +\textbf{v}.\nabla\textbf{v})=\mu\Delta\textbf{v}-\nabla p-\rho\textbf{g}\\
\end{equation}
where $\textbf{g}$ is the acceleration of gravity.

\subsection{\label{sec:levelb}BOUNDARY CONDITIONS}

At the liquid-gas interface, $z=h(r,t)$, the kinematic condition is used to relate the velocity field to interface height.
\begin{equation}\label{eq:3}
w = \frac{\partial h}{\partial t} + u\frac{\partial h}{\partial r}   
\end{equation} 
The normal and tangential component of interfacial stress balance is as follows: 
\begin{equation}\label{eq:4}
\hat{\textbf{n}}.\textbf{T}.\hat{\textbf{n}}= 2H\sigma-\Pi, \;\hat{\textbf{t}}.\textbf{T}.\hat{\textbf{n}}= t.\nabla\sigma
\end{equation}
where $\textbf{T}=\mu(\nabla u + \nabla u^T)$ is the viscous stress of the liquid, $\hat{\textbf{n}}$ and $\hat{\textbf{t}}$ are unit vector normal and tangential to the liquid-gas interface respectively which are defined as follows:
\begin{equation}\label{eq:5}
\hat{\textbf{n}}=\frac{(-\frac{\partial h}{\partial r},1)}{\sqrt{1+(\frac{\partial h}{\partial r})^2}}, \hat{\textbf{t}}=\frac{(1,\frac{\partial h}{\partial r})}{\sqrt{1+(\frac{\partial h}{\partial r})^2}}\\
\end{equation}
The curvature of the substrate is given by 
\begin{equation}\label{eq:6}
2H=-({\frac{\partial^2 h}{\partial r^2}}+\frac{1}{r}\frac{\partial h}{\partial r})\\
\end{equation}
A two term disjoining pressure, $\Pi$ \cite{Alleborn2004,Schwartz2005} which is used to relieve the stress singularity at the contact line is given by
\begin{equation}\label{eq:7}
\Pi=A\Big[(\frac{b}{h})^n- (\frac{b}{h})^m\Big]\\
\end{equation}
\\
where $A\geq0$ is the disjoining pressure parameter and exponents, $(n,m)=(3,2)$ are constant and $b$ is the precursor film thickness. The value of $A$ is related to the equilibrium contact angle, $\theta_{eq}$ as follows:\cite{Schwartz2005}
\begin{equation}\label{eq:8}
A=\frac{(n-1)(m-1)}{b(n-m)}\sigma(1-cos\theta_{eq})\\
\end{equation}
At the liquid-substrate interface, $z=0$, the imbibition condition is used:
\begin{equation}\label{eq:9}
w=\frac{\kappa}{\mu}p\\
\end{equation}
where $\kappa$ is the proportionality constant with units of length and depends on pore diameter, number of pores per unit area, and substrate thickness. \cite{Davis1999} No-slip condition is as follows: 
\begin{equation}\label{eq:10}
u=0\\
\end{equation}
The following conditions are imposed for axi-symmetric spreading:
\begin{equation}\label{eq:11}
\frac{\partial h}{\partial r}\bigg|_{r=0}=0, 
\frac{\partial^3h}{\partial r^3}\bigg|_{r=0}=0
\end{equation} 
For droplet volume $V_0$ to be conserved, following equation is used:
\begin{equation}\label{eq:12}
\int\limits_{0}^{r_0} 2\pi rh(r)dr=V_0+\pi b
\end{equation}
where subscript 0 denotes the initial value of a quantity and $\pi b$ is the correction factor for precursor film on a circular region with radius equal to the characteristic value.

\subsection{\label{sec:levelc}SCALINGS}
The governing equations and the corresponding boundary conditions are normalized using following factors:\\
\begin{equation}\label{eq:13}
r=l\tilde r, z=h\tilde z, u=u_{sc}\tilde u, w=\epsilon u_{sc}\tilde w$$
$$
t=\frac{l}{u_{sc}}\tilde t, p=p_{sc}\tilde p, \Pi=p_{sc}\tilde \Pi
\end{equation} 
\\
where $u_{sc}=\frac{\epsilon^{3}\sigma}{3\mu}$ is the characteristic spreading speed, $p_{sc}=\frac{h\sigma}{l^{2}}$ is the pressure scale, $b=h\tilde b$ is the precursor film thickness, and $\epsilon=\frac{h}{l}$ is the lubrication ratio. The effective permeability of the substrate is given by $\kappa=\frac{dh^{3}}{3l^{2}}\tilde\kappa$ where $d$ is the thickness of the substrate. The non-dimensional quantities are denoted with tilde. The droplet is assumed to be thin, which means the ratio of characteristic height, $h$, to characteristic radius, $l$, is much less than unity ($\epsilon\ll1$). Hence, the lubrication approximation is applied which reduces the governing equations and boundary conditions to:\\
Continuity equation:
\begin{equation}\label{eq:14}
\frac{\partial {\tilde w}}{\partial {\tilde z}}+\frac{1}{\tilde r}\frac{\partial({\tilde r \tilde u})}{\partial \tilde r}=0,\\
\end{equation}
r-momentum equation:
\begin{equation}\label{eq:15}
\frac{\partial^2{\tilde u}}{\partial{\tilde z}^2}-3\frac{\partial{\tilde p}}{\partial {\tilde r}}=0, 
\end{equation}
z-momentum equation:
\begin{equation}\label{eq:16}
-\frac{\partial{\tilde p}}{\partial{\tilde z}}-B=0
\end{equation}
where $B=\frac{l^{2}\rho g}{\sigma}$ is the Bond number.\\
The kinematic condition at $\tilde z=\tilde h(\tilde r,\tilde t)$,  is as follows:
\begin{equation}\label{eq:17}
\tilde w = \frac{\partial \tilde h}{\partial \tilde t} + u\frac{\partial \tilde h}{\partial \tilde r},
\end{equation}
The normal and tangential component of interfacial stress balance is written as follows:
\begin{equation}\label{eq:18}
\tilde p=-(\frac{\partial^2{\tilde h}}{\partial{\tilde r^2}}+\frac{1}{\tilde r}\frac{\partial{\tilde h}}{\partial{\tilde r}})-\tilde A\Big[(\frac{\tilde b}{\tilde h})^{3}-(\frac{\tilde b}{\tilde h})^{2}\Big], 
\end{equation} 
\begin{equation}\label{eq:19}
\frac{\partial{\tilde u}}{\partial{\tilde z}}=0\\
\end{equation}
where $\tilde A=\frac{A}{p_{sc}}$.\\
Imbibition condition at $\tilde z=\tilde b$ is as follows:
\begin{equation}\label{eq:20}
\tilde w=-\frac{\tilde \kappa}{\mu d}\tilde p\\
\end{equation}
No-slip condition at $\tilde z=0$ is as follows:
\begin{equation}\label{eq:21}
\tilde u=0
\end{equation}

\subsection{\label{sec:leveld}NUMERICAL SCHEME}

The non-dimensional time-evolution equation for the droplet height is obtained by integrating eqns.(\ref{eq:14}-\ref{eq:16}) subject to the boundary conditions (\ref{eq:17}-\ref{eq:21}):

\begin{equation}\label{eq:22}
\frac{\partial{\tilde h}}{\partial{\tilde t}}=\frac{1}{\tilde r}\frac{\partial}{\partial{\tilde r}}\Bigg\{\tilde r\tilde h^3\frac{\partial}{\partial{\tilde r}}\Bigg(B\tilde h-\tilde A{\tilde{\Pi}}-\frac{\partial^2 \tilde h}{\partial \tilde r^2}-\frac{1}{\tilde r}\frac{\partial{\tilde h}}{\partial{\tilde r}}\Bigg)\Bigg\}
-\tilde \kappa\Bigg\{B(\tilde h-\tilde b)-\tilde A{\tilde \Pi}-\frac{\partial^2 \tilde h}{\partial \tilde r^2}-\frac{1}{\tilde r}\frac{\partial{\tilde h}}{\partial{\tilde r}}\Bigg\}
\end{equation}
The solution of eqn.(\ref{eq:22}) is subject to the following boundary conditions on the domain 
$0\leq\tilde r\leq\textit{L}$:
\begin{equation}\label{eq:23}
\frac{\partial{\tilde h}}{\partial{\tilde r}}(0,\tilde t)=0,\tilde h(L,\tilde t)=\tilde b,
\frac{\partial^3 h}{\partial{\tilde r^3}}(0,\tilde t)=0,\frac{\partial{\tilde h}}{\partial{\tilde r}}(L,\tilde t)=0.
\end{equation}

The initial droplet profile is obtained in the form of a fourth-degree polynomial satisfying the above boundary conditions (\ref{eq:23}) and the volume constraint (\ref{eq:12}). Computational domain length, \textit{L} is considered to be 4 with 600-700 points per unit length. The spatial derivatives are discretised using second-order central finite difference scheme  which converts the PDE into sets of differential algebraic equations (DAE). For the resulting system of DAEs, an adaptive time-step solver, DASSL, is used which is available in SLATEC library. 

The apparent contact angle is calculated by finding maximum slope of the interface and multiplying by $\epsilon$ to convert into dimensional form. The radius is obtained by finding the distance between the z-axis and the intersection point of the apparent contact line with the r-axis.

\subsection{\label{sec:levelc}NORMALISING FACTORS}

In order to obtain eqn.(\ref{eq:22}), two normalising factors are used. The first one is based on the radius and height of the droplet at the onset of equilibrium and the second one is described by those values at the initial stage. 
Based on this definition, the required parameters are calculated using both normalising factors as described below:\\
a. \underline{First normalising factor}:
Using equilibrium values ($l_{eq}$,$h_{eq}$) as normalising factors, the following input parameters are obtained: Normalised initial radius, $r_0=\frac{l_{in}}{l_{eq}}$;
normalised initial height,
$h_0=\frac{h_{in}}{h_{eq}}$; initial volume based on spherical-cap approximation,
$V_{in}=\frac{\pi h_{in} (3l_{in}^{2}+ h_{in}^{2})}{6}$; final volume when the droplet ceases to exist on the substrate (assuming spherical-cap),
$V_{eq}=\frac{\pi h_{eq} (3l_{eq}^{2}+ h_{eq}^{2})}{6}$;
normalised initial volume,
$V_0=\frac{V_{in}}{V_{eq}}$;
Bond number,
$B_{eq}=\frac{l_{eq}^{2} \rho g}{\sigma}$;
time
$t_{eq}=\frac{h_{eq}^{3}\sigma}{3l_{eq}^{4}\mu}\tilde t$;
and lubrication ratio,
$\epsilon_{eq}=\frac{h_{eq}}{l_{eq}}$.
\\
b. \underline{Second normalising factor}:
Using initial values ($l_{in}$,$h_{in}$) as normalising factors, the following input parameters are found as:
$r_0=\frac{l_{in}}{l_{in}}=1$;
$h_0=\frac{h_{in}}{h_{in}}=1$;
$V_0=\frac{V_{in}}{V_{in}}=1$;
$B_{in}=\frac{l_{in}^{2} \rho g}{\sigma}$;
$t_{in}=\frac{h_{in}^{3}\sigma}{3l_{in}^{4}\mu}\tilde t$;
and $\epsilon_{in}=\frac{h_{in}}{l_{in}}$

\section{\label{sec:level4}RESULTS AND DISCUSSION}

For a small droplet, where the effect of gravity is negligible, the driving force for spreading is capillary pressure. The maximum extent of spreading is dictated by the dominant effect of disjoining pressure over capillary pressure in the later stages. For a smooth, impermeable substrate, a perfectly wetting Newtonian liquid obeys Tanner's law.\cite{Tanner1979} However, the maximum spreading radius will vary if the liquid partially wets the substrate. Under the combined effect of attractive and repulsive components of intermolecular forces in disjoining pressure, a liquid partially wets the substrate and thus, has a finite non-zero equilibrium contact angle.\cite{Glasner2003} For smooth permeable substrates, the spreading occurs at two stages: First, the inertial stage where lubrication ratio is much larger than unity. At this stage, the droplet spreads to a maximum radius and stops momentarily. Second, the equilibrium stage which is signified by a decrease in droplet volume due to the imbibition. As a result, the droplet radius decreases and the contact angle remains constant over a particular time before becoming zero.
At the equilibrium stage, the lubrication ratio is much less than unity.\\ 
A 3-step procedure is followed in order to make comparison of model predictions with experimental data:\\
(i) A specific value of $b$ which yields $\tilde b$ through the scaling is used as a starting point. \\
(ii) Normalised initial radius ($r_0$) and height ($h_0$) are obtained using both normalising factors. These parameters along with $\tilde b$ are utilised to obtain the initial droplet profile for each case. The model parameters used for matching with the experimental values are disjoining pressure parameter, $\tilde A$ and non-dimensional imbibition coefficient, $\tilde k$. $\tilde A$ governs the equilibrium stage of spreading and its higher values signifies less hydrophilicity or higher equilibrium contact angle. The equilibrium stage is first matched by varying $\tilde A$, then $\tilde k$ is adjusted so as to match the imbibition time. Many combinations of ($\tilde A$, $\tilde k$) are tried for a specific $\tilde b$ until good agreement with the experimental results is found.\\
(iii) For both impermeable and permeable substrates, ($A, k$) corresponding to each value of $b$ are obtained using the first normalising factor. The simulation is then conducted using these values and the second normalising factor. This approach provides a clear picture on the role of each normalising factor on the numerical predictions.

\begin{figure}
\includegraphics[scale=0.35]{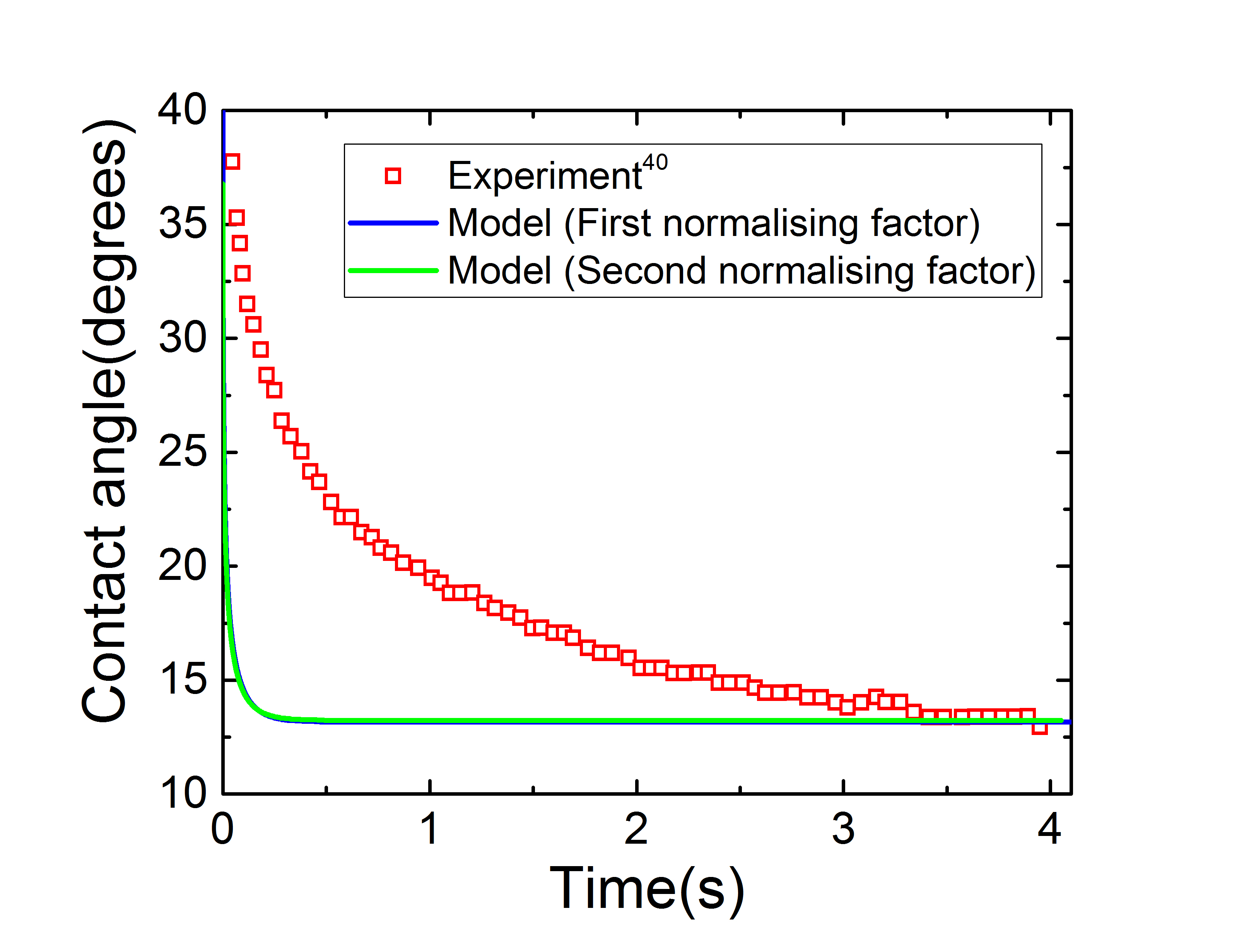}
\caption{Comparison between experimental data and simulation results of contact angle of squalene spreading on glass.\cite{Ruijter1997} The simulation parameters used are: $V_{in}=V_{eq}=6.046\;\mu$L, $l_{in}=1.6$ mm, $l_{eq}=2.548$ mm, 
$\epsilon_{in}=0.781$, $\epsilon_{eq}=0.229$, $B_{in}=0.702$, $B_{eq}=1.78$, $b=0.98\;\mu$m, $A=8271\;$N$m^{-2}$}
\label{fig:two}
\end{figure}

In Fig.~\ref{fig:two}, the contact angle of squalene drop spreading on glass \cite{Ruijter1997} is compared with the model predictions. There is no change in volume over time since glass is impermeable. The spherical-cap approximation is used for calculating the volume of the droplet. First the initial values are normalised with the equilibrium characteristic values. For a specific $b$ ($<$1 $\mu$m \cite{Popescu2012}), the initial shape of the droplet is obtained. As can be seen in Fig.~\ref{fig:two}, there is a good agreement between model predictions and experimental data, only at the equilibrium stage of spreading and the model failed in predicting the earlier stages of spreading. Then, using the corresponding ($b,A$) obtained by the first normalising factor, the simulation is performed based on the second normalising factor. A clear overlap in the numerical prediction highlights the fact that either normalising factor can predict the equilibrium contact angle. The model was further tested for $b=0.75\;\mu$m and $0.5 \;\mu$m and a good match was observed (The results are not provided since the behaviour was exactly the same). For each $b$, the numerical predictions coincided using both normalising factors.

\begin{figure}[h]
\includegraphics[scale=0.35]{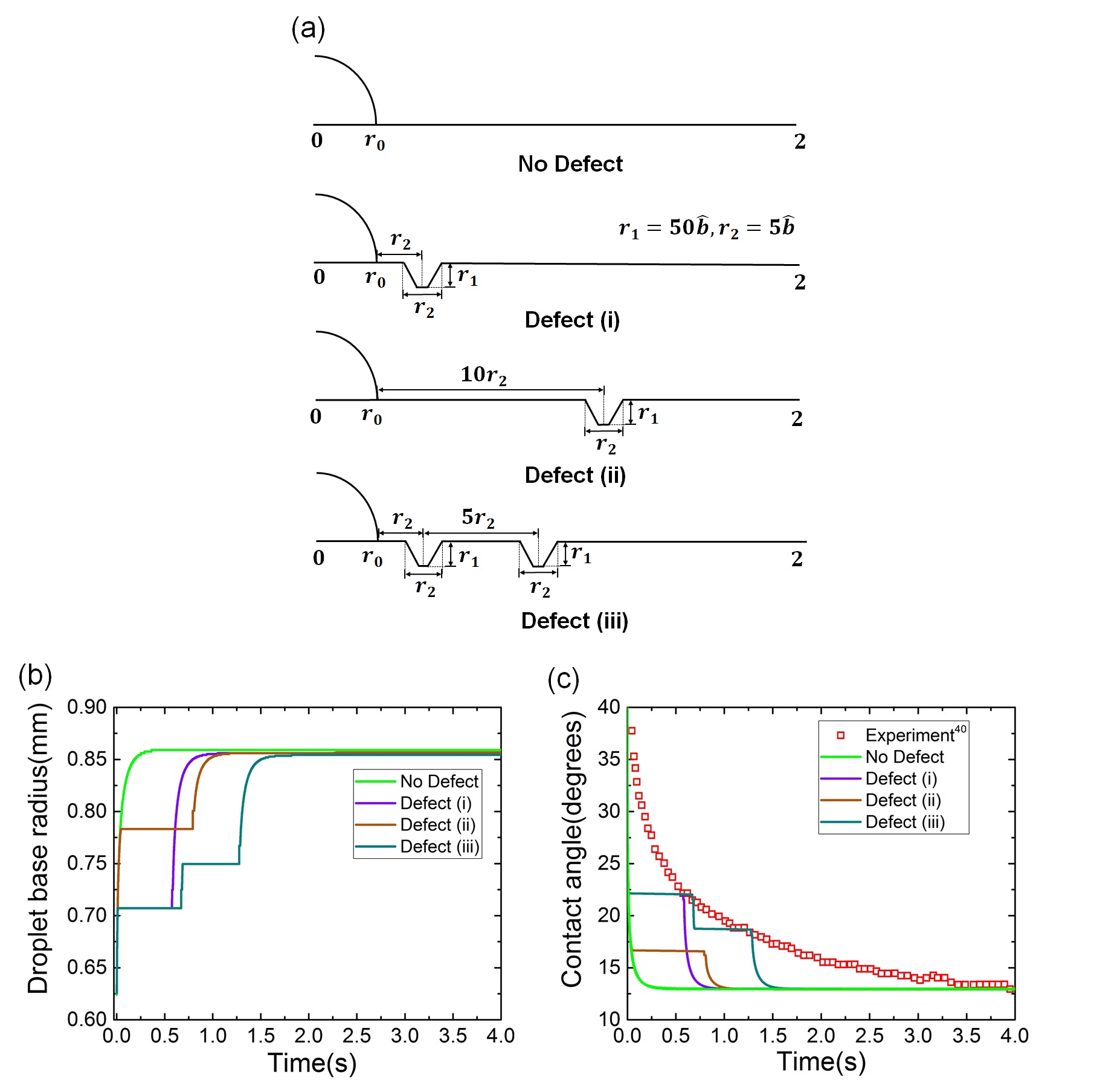}
\caption{(a) The nature of defects considered. Comparison between experimental data and simulation results of (b) droplet base radius, and (c) contact angle of squalene spreading on glass in the presence of substrate roughness. \cite{Ruijter1997} The simulation parameters remain unchanged.}
\label{fig:three}
\end{figure}

The decrease in contact angle is sharper in numerical simulation compared to the experiments and equilibrium is reached earlier. The difference between numerically predicted results and experiments might be attributed to the substrate roughness. Since any substrate possesses inherent roughness (unless otherwise specified), one way to delay the onset of equilibrium is to take into account the effect of substrate roughness in the mathematical model.  In the presence of roughness, the contact line is pinned and the apparent contact angle increases.\cite{Lafuma2003}
To incorporate roughness in the model, a shape function, $\tilde z=\gamma(\tilde r)$ is introduced in this study. The governing equations remain unchanged in the presence of substrate roughness. However, the kinematic condition and the interfacial stress balance components are evaluated at $\tilde z=\gamma(\tilde r)+\tilde h(\tilde r,\tilde t)$. The imbibition and no-slip conditions also occur at $\tilde z=\gamma(\tilde r)+\tilde b$ and at $\tilde z=\gamma(\tilde r)$ respectively. Based on these modifications, the droplet evolution equation takes the following form:

\begin{eqnarray}       
\frac{\partial{\tilde h}}{\partial{\tilde t}}=\frac{1}{\tilde r}\frac{\partial}{\partial{\tilde r}}\Bigg\{\tilde r\tilde h^3\frac{\partial}{\partial{\tilde r}}\Bigg(B(\tilde h+\gamma)-\tilde A{\tilde{\Pi}}-\frac{\partial^2 (\tilde h+\gamma)}{\partial \tilde r^2}-\frac{1}{\tilde r}\frac{\partial{(\tilde h+\gamma)}}{\partial{\tilde r}}\Bigg)\Bigg\} \nonumber \\
- \tilde \kappa\Bigg\{B(\tilde h-\tilde b)-\tilde A{\tilde \Pi}-\frac{\partial^2 (\tilde h+\gamma)}{\partial \tilde r^2}-\frac{1}{\tilde r}\frac{\partial{(\tilde h + \gamma)}}{\partial{\tilde r}}\Bigg\} \label{eq:24} 
\end{eqnarray}              

There are complicated forms of shape function, $\gamma(r)$, in the literature that are able to imitate a real rough substrate very accurately.\cite{Chow1998,Savva2010} However, in this study, the most simplified form of this function, known as Gaussian model,\cite{Herminghaus2012} is utilised.
The Gaussian model of roughness is represented by the following equation:

\begin{equation}\label{eq:25}
\gamma(r)=-r_1exp\Bigg(-\frac{(r-r_{loc})^2}{2r_2^2}\Bigg)
\end{equation}
where $r_1,r_2=O(\tilde b)$ and $r_{loc}$ represents the location of the defect.

For the present simulation, $r_1$ and $r_2$ are selected as $50~\tilde b$ and $5~\tilde b$, respectively. To investigate the role of defects on contact line motion, three different configurations are selected, as shown in Fig.~\ref{fig:three}(a). When the defect is placed very close to the initial shape (Defect~(i) in Fig.~\ref{fig:three}), the pinning phenomenon is observed which is represented by no change in radius over a certain time period. If the nature of the defect remains unchanged but is placed much farther from the initial shape (Defect~(ii)), pinning persists for comparatively longer period. This is attributed to non-uniform speed of droplet spreading. The characteristic speed of a spreading droplet is much higher at the initial stages and as time passes the spreading front slows down owing to substrate irregularities. Since the height of the droplet also decreases over time, the rate of conversion of the potential energy (owing to greater height in the initial stages) to kinetic energy diminishes which is manifested in the speed of the spreading front. As a result, for a defect placed much farther on the path, the contact line has a tendency to stay pinned for a longer time and thus, the equilibrium is delayed. If there is another identical defect placed at a certain distance from the first defect (Defect~(iii)), a staircase-like behaviour is observed marked by a delayed equilibrium. Our preliminary investigations showed that this staircase behaviour depends on the nature, density and spacing of defects. The detailed study of staircase bahviour with the aim of acquiring more realistic replication of a rough substrate is the topic of our ongoing research. 

\begin{figure}
\includegraphics[scale=0.35]{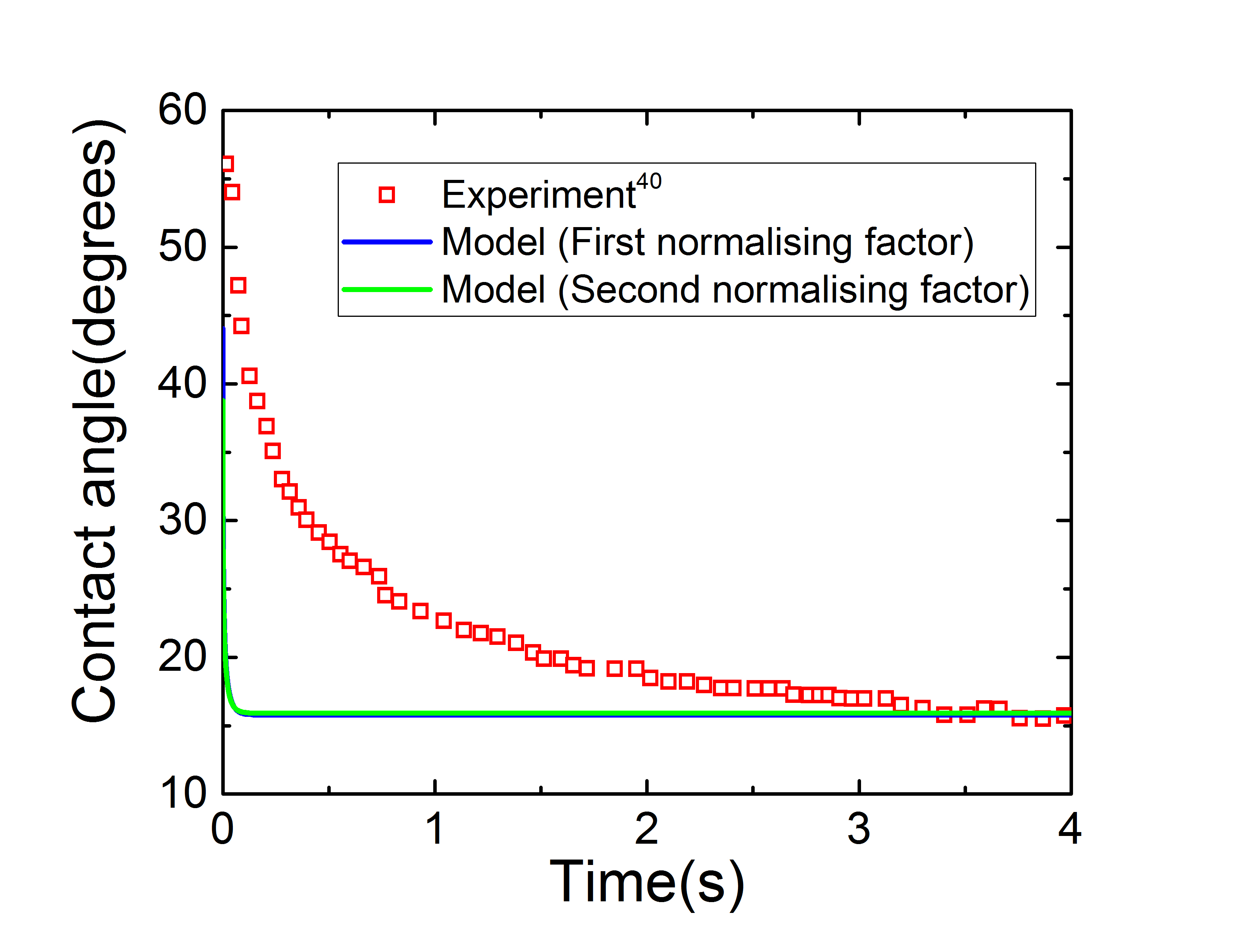}
\caption{Comparison between experimental data and simulation results of contact angle of DBP spreading on PET.\cite{Ruijter1997} The simulation parameters used are: $V_{in}=V_{eq}=5\;\mu$L, $l_{in}=1.454$ mm, $l_{eq}=2.238$ mm, 
$\epsilon_{in}=0.839$, $\epsilon_{eq}=0.277$, $B_{in}=0.635$, $B_{eq}=1.504$, $b=0.98\;\mu$m, $A=13587.2\;$N$m^{-2}$ }
\label{fig:four}
\end{figure}

In another study of impermeable substrate, the contact angle of di-n-butyl phthalate (DBP) spreading on poly(ethyleneterephthalate) (PET) is considered.\cite{Ruijter1997} A similar procedure as the previous case is followed and a good agreement with the experimental data is achieved for $b=0.98\;\mu$m (Fig.~\ref{fig:four}), $0.75\;\mu$m, and $0.5\;\mu$m. Comparing the model predicted results and experiments show that irrespective to the applied characteristic lengths (for radius and height) and precursor film, $b$, the equilibrium contact angle is mainly governed by the proper choice of $A$.  

\begin{figure}
\includegraphics[scale=0.35]{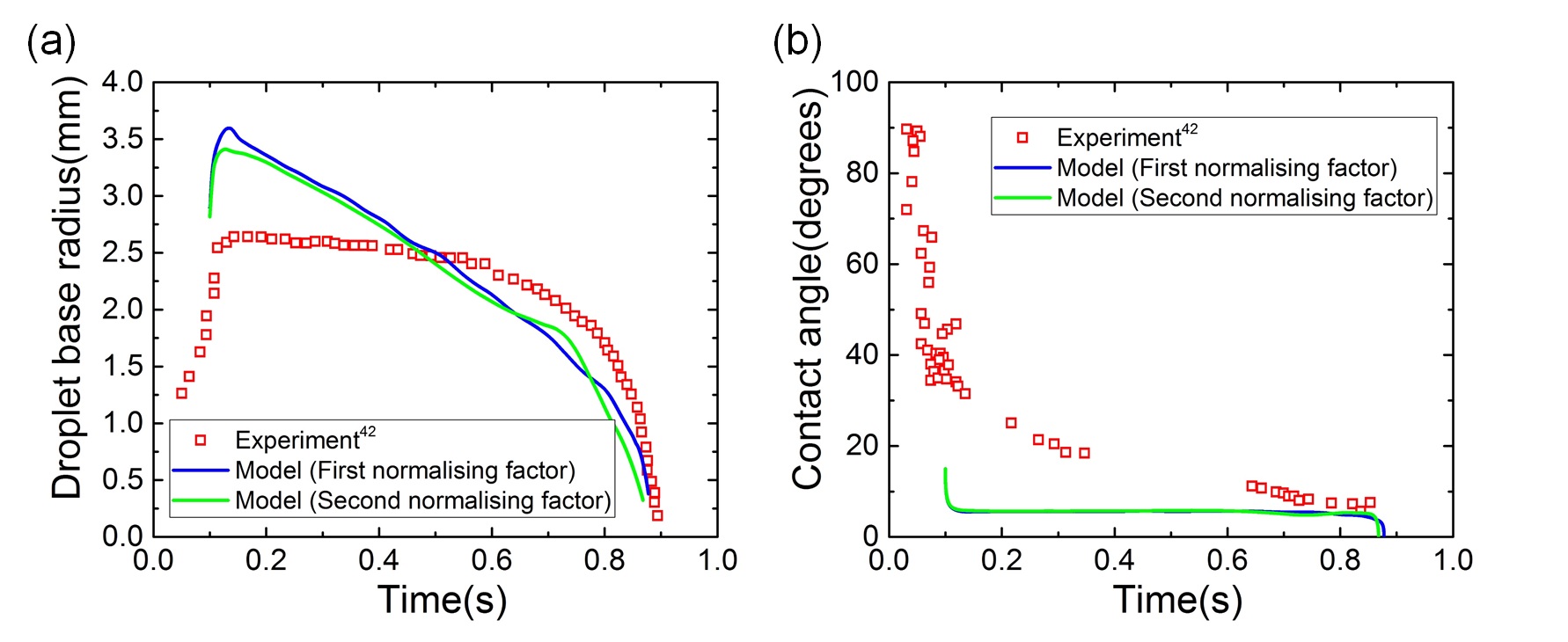}
\caption{Comparison between experimental data and simulation results of (a) droplet base radius, and (b) contact angle of water droplet spreading on a h-PBT-4 mat.\cite{Wang2016} The simulation parameters used are: $V_{in}=10.04\;\mu$L, $V_{eq}=1.089\;\mu$L, $l_{eq}=2.81$ mm, $l_{in}=2.087$ mm, 
$\epsilon_{in}=0.281$, $\epsilon_{eq}=0.076$, $B_{in}=1.076$, $B_{eq}=0.594$, $b=0.98\;\mu$m, $A=6379.2\;$N$m^{-2}$, $k/d=5.076\times10^{-8}$}
\label{fig:five}
\end{figure}

For a permeable substrate, model predictions and experimental measurements of water droplet spreading on a poly(butylene ter-ephthalate) (h-PBT-4) mat are compared in the absence of contact line pinning for $b=0.98\;\mu$m 
(Fig.~\ref{fig:five}). In the experimental work, the onset of lubrication phase was reported as $t=0.117\;$s. The corresponding initial height ($h_{in}$) and radius  ($l_{in}$) of the droplet were $0.79\;$mm and $2.81\;$mm, respectively. The equilibrium height ($h_{eq}$) and radius ($l_{eq}$) were reported to be $0.159\;$mm and $2.087\;$mm, respectively. The initial droplet profile is obtained based on these values. As can be observed in Fig.~\ref{fig:five}, there is an overlap between numerical predictions using either normalising factor for constant ($b,A, k$). In both cases, droplet radius increases at the initial stages followed by a retraction due to permeation which is similar to what is observed in experiments. Using other $b$ values ($b=0.75\;\mu$m and $0.5\;\mu$m), the same results are obtained. This reinforces our hypothesis that using either normalising factors enables accurate prediction of the experimental values, for both permeable and non-permeable substrates.\\
The time lag between the numerical prediction and experimental values is primarily related to the starting point of simulation at $t=0.117\;$s as this marks the onset of lubrication. A possible explanation might be the difference between experimental and numerical initial droplet profile to run the simulation. 

\begin{figure}
\includegraphics[scale=0.35]{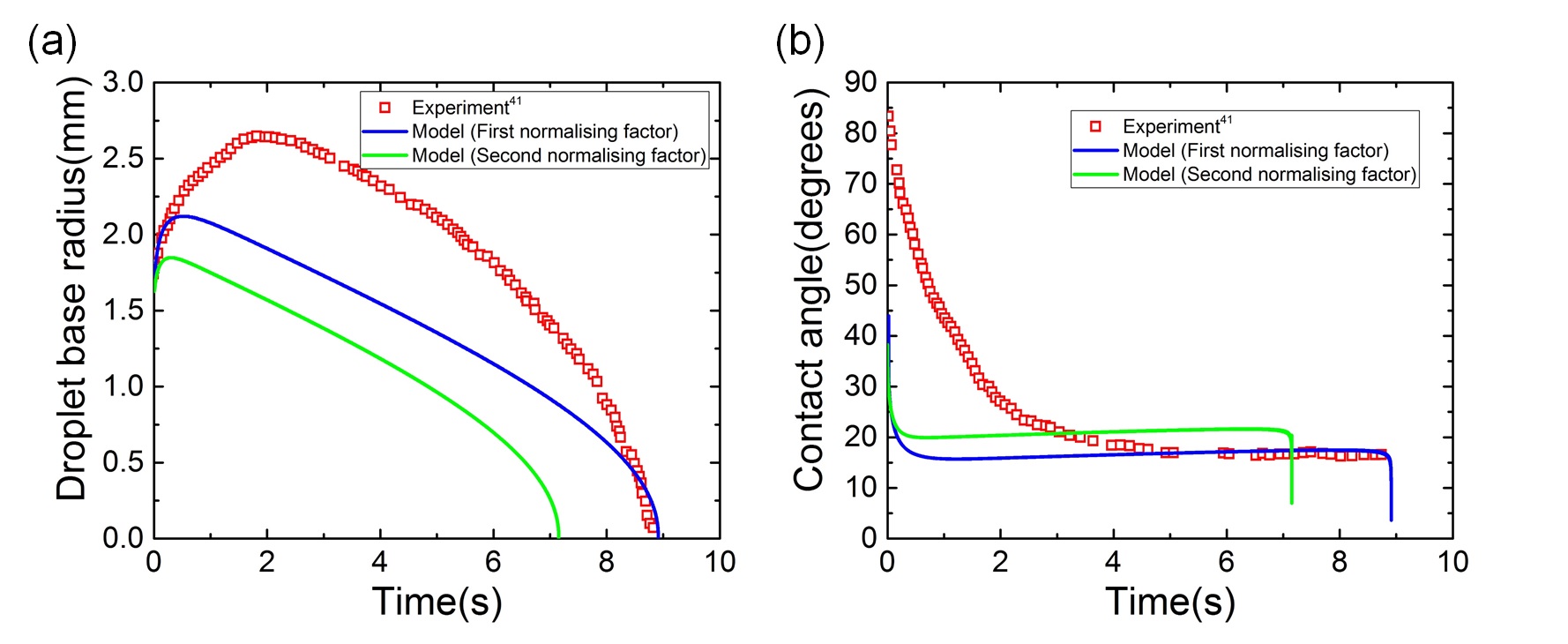}
\caption{Comparison between experimental data and simulation results of (a) droplet base radius, and (b) contact angle of SO300 droplet spreading on UDMAT.\cite{Kumar2006} The simulation parameters used are: $V_{in}=7.788\;\mu$L, $V_{eq}=1.609\;\mu$L, $l_{in}=1.7$ mm, $l_{eq}=2.32$ mm, $\epsilon_{in}=0.82$, $\epsilon_{eq}=0.082$, $B_{in}=1.404$, $B_{eq}=2.615$, $b=0.98\;\mu$m, $A=9633$ N$m^{-2}$, $k/d=2.039\times10^{-6}$} 
\label{fig:six}
\end{figure}

To showcase another example of the spreading phenomena on a permeable substrate, the droplet radius and contact angle of silicone oil grade SO300 spreading on unidirectional mat (UDMAT) \cite{Kumar2006} are compared with the model predictions. The initial and equilibrium height and radius of the droplet were obtained at $t=0$ and at the onset of equilibrium stage respectively. Using first normalising factor resulted in better agreement with experimental data for $b=0.98\;\mu$m (Fig.~\ref{fig:six}) and $0.75\;\mu$m. When compared to the second normalising factor, there is an offset in prediction which might be due to the droplet profile employed in this case for running simulation. 

The significant difference between model prediction of the contact angle (Fig.~{\ref{fig:six}}(b)) for $t<2\;$s is attributed to large lubrication ratio ($\epsilon\gg1$) at the inertial stage. The second normalising factor hinges on inertial values which was not measured accurately. Even though dynamic radius and height measurements are exclusively reported, the literature did not provide any proper explanation for the calculation of the contact angle.\cite{Kumar2006}

\section{\label{sec:level4}MODIFIED SCALING FOR BETTER PREDICTION OF EQUILIBRIUM CONTACT ANGLE}

Theoretically, the disjoining pressure parameter, $A_{th}$ is related to the equilibrium contact angle, $\theta_{eq}$ as follows:\cite{Schwartz2005}

\begin{equation} \label{eq:26}
A_{th}=\frac{(n-1)(m-1)}{b(n-m)}\sigma(1-cos\theta_{eq})\;or,\frac{2}{b}\sigma(1-cos\theta_{eq})\\
\end{equation}

To numerically model the spreading process, it is essential to start with an initial guess of $\tilde A$. Since the equilibrium contact angle, $\theta_{eq}$, is known from the dynamic contact angle measurement, it is possible to have an estimate of $\tilde A$ ($A=p_{sc}\tilde A$). 

It is found that the suggested $A$ (from eqn.26) cannot predict the equilibrium contact angle accurately. Hence, the following modification is proposed: 
\begin{equation}\label{eq:25}
A_{m}=\frac{A_{th}}{\epsilon^n}
\end{equation} where $n$ is the exponent to be ascertained for impermeable and permeable substrates.

\begin{figure}
\includegraphics[scale=0.35]{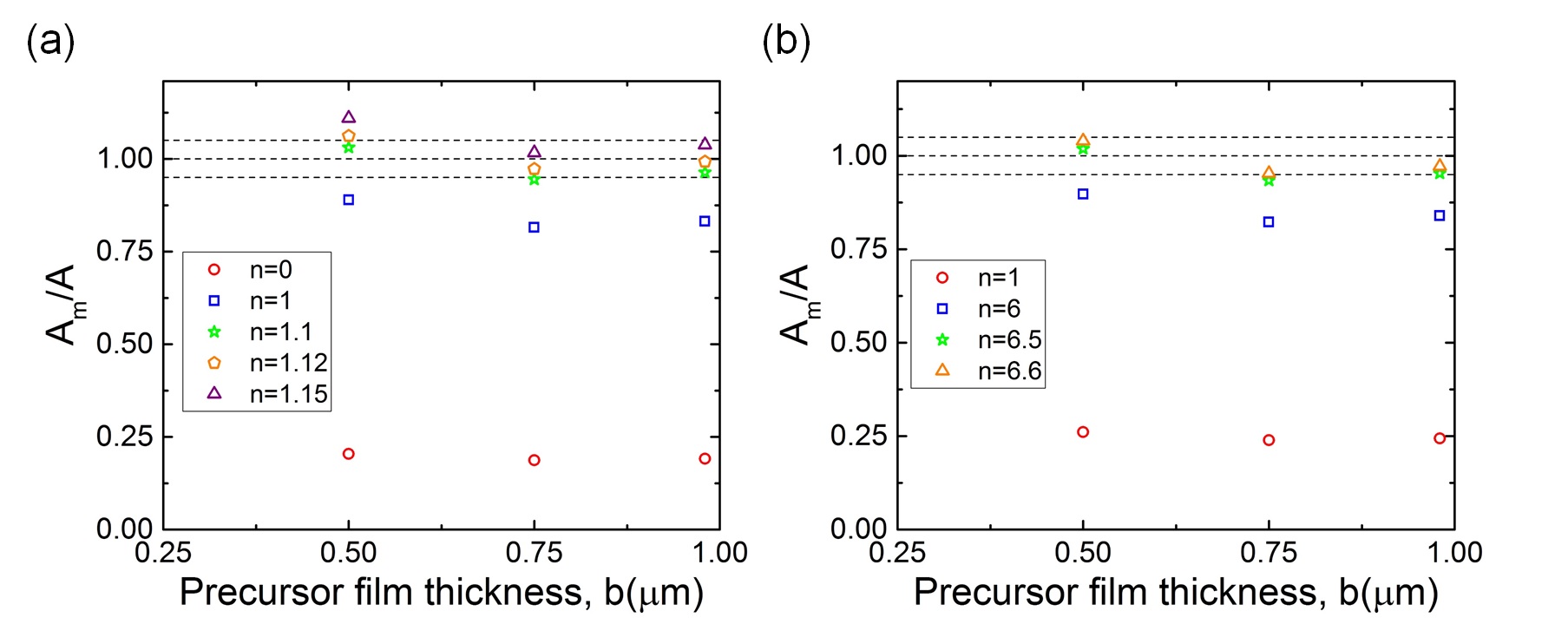}
\caption{Ratio $A_{m}/A$ for (a) First normalising factor and (b) Second normalising factor of  squalene spreading on glass with $\theta_{eq}=12.88^{\circ}$.\cite{Ruijter1997}} 
\label{fig:seven}
\end{figure}

\begin{figure}
\includegraphics[scale=0.35]{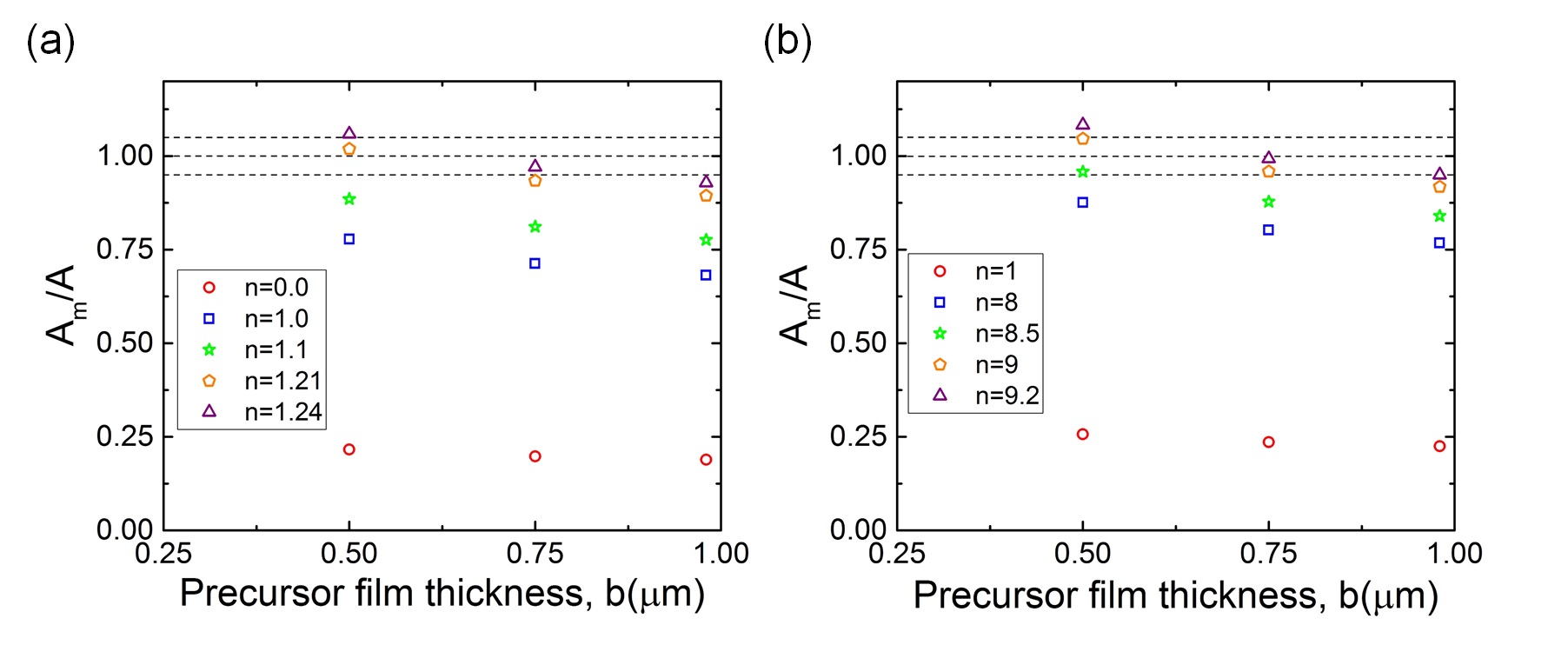}
\caption{Ratio $A_{m}/A$ for (a) First normalising factor and (b) Second normalising factor of DBP spreading on PET with $\theta_{eq}=15.569^{\circ}$.\cite{Ruijter1997}}  
\label{fig:eight}
\end{figure}

The role of both normalising factors on the prediction of $n$ for droplet spreading on impermeable and permeable substrate is presented next.

a. \underline{First normalising factor}:

(i) Impermeable cases: The squalene droplet spreading on glass is considered as an impermeable case. Starting with the first normalising factor, $n$ is varied from 0 to 1.15 until the ratio $A_{m}/A$ was close to 1, with an error of $\pm$5$\%$. According to Fig.~\ref{fig:seven}(a), choosing $n=1.12$ resulted in a better starting guess of $\tilde A$ for a particular precursor film thickness. Similarly, for DBP spreading on PET, $n$ is changed from 0 to 1.3 (Fig.~\ref{fig:eight}(a)). A choice of $n=1.24$ has led to a better starting guess of $\tilde A$ for each value of precursor film thickness.

\begin{figure}
\includegraphics[scale=0.35]{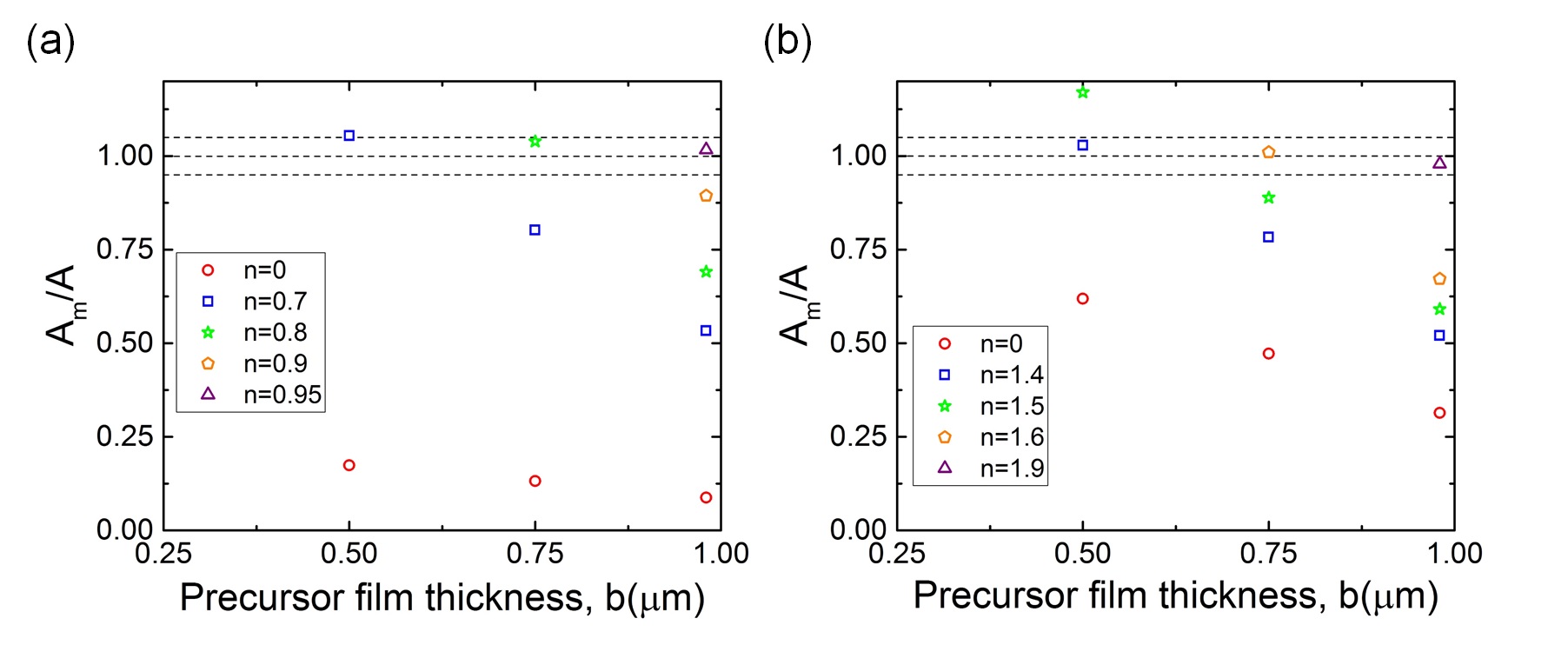}
\caption{Ratio $A_{m}/A$ for (a) First normalising factor and (b) Second normalising factor of water droplet spreading on a h-PBT4 mat with $\theta_{eq}=5^{\circ}$. \cite{Wang2016}}
\label{fig:nine}
\end{figure}

\begin{figure}
\includegraphics[scale=0.35]{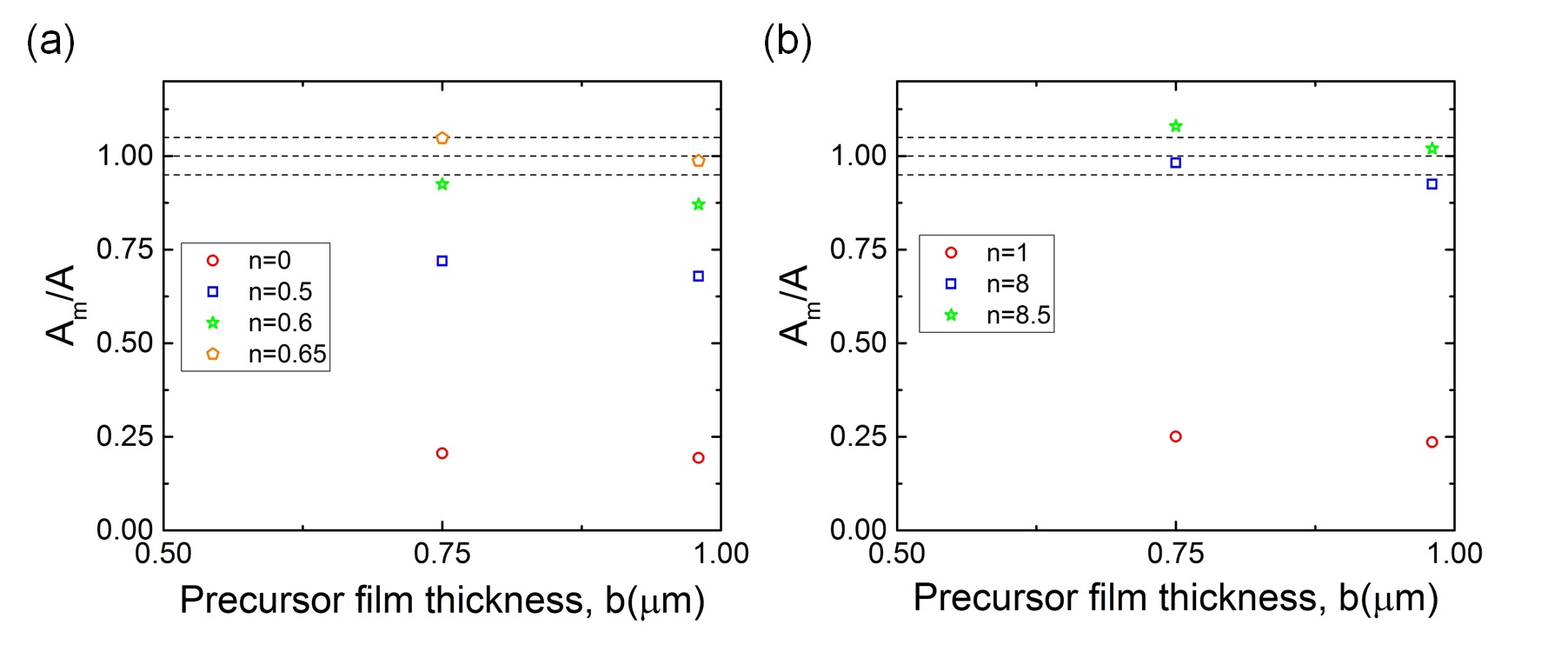}
\caption{Ratio $A_{m}/A$ for (a) First normalising factor and (b) Second normalising factor of SO300 spreading on UDMAT with $\theta_{eq}=17^{\circ}$.\cite{Kumar2006}}
\label{fig:ten}
\end{figure}

(ii) Permeable cases: For water droplet spreading on h-PBT-4 mat, $n$ is obtained to be $0.95$, $0.8$, $0.7$ for $b=0.98\;\mu$m, $0.75\;\mu$m, and $0.5\;\mu$m, respectively (Fig.~\ref{fig:nine}(a)). Following the similar procedure for SO300 droplet spreading on UDMAT, the value of $n$ is evaluated to be $0.65$ with an error within $\pm$5$\%$ for $b=0.98\;\mu$m and $0.75\;\mu$m (Fig.~\ref{fig:ten}(a)).

b. \underline{Second normalising factor}:

(i) Impermeable cases: Figs.~\ref{fig:seven}(b) and ~\ref{fig:eight}(b) show that, using second normalising factor the value of $n$ are obtained to be $6.6$ and $9.2$ for squalene droplet spreading on glass and DBP spreading on PET, respectively. It is noteworthy that in both experiments, no effort was made to precisely track the onset of lubrication. This could be one plausible explanation for such a big variation in the value of $n$ for these impermeable cases.

(ii) Permeable cases: For water droplet spreading on h-PBT-4 mat, the values of $n=1.9$, $1.6$, and $1.4$ are obtained for $b=0.98\;\mu$m, $0.75\;\mu$m, and $0.5\;\mu$m, respectively (Fig.~\ref{fig:nine}(b)). The onset of lubrication was precisely tracked. The second normalising factor yields value of $n=8.5$ for $b=0.98$ and $0.75\;\mu$m for SO300 spreading on UDMAT. 

In summary, first normalising factor yields $n$ between 1.0-1.3 for impermeable and 0.5-1.0 for permeable substrates, thereby leading to a better starting guess for $\tilde A$. For second normalising factor, it is essential to precisely track the initial conditions. Only one experimental study on a permeable substrate meets this criterion.\cite{Wang2016} Hence, in the absence of profound evidence, it is difficult to specify a range of $n$ to be used for the second normalising factor.

\section{\label{sec:level5}CONCLUSION}

Ability of a mathematical model to accurately predict the droplet radius and apparent contact angle is governed by the choice of radius and height at the initial and equilibrium stages of spreading. Using four examples of droplet spreading over impermeable and permeable substrates, it is observed that choosing radius and height in either stage of the spreading process provides similar prediction of the experimentally obtained base radius and equilibrium contact angles. In addition, incorporating the substrate roughness into the model is found to bridge the gap between model predictions and experiments for initial stages of spreading. A modified scaling which relates the theoretical value of the disjoining pressure parameter to its numerical counterpart through the lubrication ratio is also proposed. The lubrication ratio takes different exponents based on the permeability of the substrate. For impermeable and permeable substrates, $n$ was found to be in the range of 1.0-1.3 and 0.5-1.0 respectively. The proposed method greatly simplified the initial guess for the disjoining pressure parameter in the numerical simulation as previously there was no possibility of ascertaining the value. This study has shown that it is always possible to predict the value of the lubrication exponent using the first normalising factor which depends on the equilibrium condition. However, the second normalising factor depends on the initial conditions entirely and thus, without precise tracking of the lubrication ratio it is not feasible to suggest a range of $n$ for impermeable/permeable substrates. Overall, the equilibrium stage characteristic parameters is more appropriate since the modified scaling allows prediction of the numerical disjoining parameter with $\pm5\%$ error. As a future study, more effort is needed for the accurate calculation of the initial droplet profile and shape approximation for volume at each stage of spreading. The hypotheses provided in this work is expected to hold true when the  model is extended to include the whole spectrum of spreading, especially the inertial stages when the lubrication ratio is much greater than unity.

\bibliography{Choi2012}

\end{document}